\documentclass[
  aps,
  prl,
  twocolumn,
  amsfonts,
  amssymb,
  amsmath,
  superscriptaddress,
  floatfix
]{revtex4-2}

\usepackage[utf8]{inputenc}
\usepackage{bm}  
\usepackage{siunitx}
\usepackage{booktabs}
\usepackage{graphicx}
\usepackage[svgnames]{xcolor}
\usepackage{adjustbox}
\usepackage{tabularx}
\usepackage{xspace}  
\usepackage{comment}
\usepackage{tikz} 
\usepackage{import} 

\usepackage[hidelinks]{hyperref}

\newcommand*{\vb}[1]{\boldsymbol{#1}}  

\newcommand*{\vvec}{\vb{v}}

\begin{document}

\title{Kelvin wave and soliton propagation in classical viscous vortex filaments} 

\author{Elio Sterkers}
\affiliation{%
Université Côte d'Azur, Observatoire de la Côte d'Azur, CNRS,
Laboratoire Lagrange, Boulevard de l'Observatoire CS 34229 - F 06304 NICE Cedex 4, France
}
\author{Giorgio Krstulovic}
\affiliation{
Universit\'{e} C\^{o}te d'Azur, CNRS, Institut de Physique de Nice {\color{black}(INPHYNI)}, 17 rue Julien Lauprêtre, 06200 Nice, France}

\date{\today}

\begin{abstract}
  Vortex filaments are highly rotating localized structures of fluids that admits several types of excitation. Here, we study them by using numerical simulations of the three-dimensional incompressible Navier-Stokes equations. We first address the propagation of Kelvin waves, helicoidal excitations propagating along the filament, and measure their dispersion relation which turns out to be in good agreement with the original Lord Kelvin predictions. Then, inspired by the connection between vortex line dynamics and an integrable system, we show numerically the existence of solitons propagating along vortex filaments and study the collision of two of such structures. Finally, we show numerically the experimental feasibility of studying vortex solitons in the lab, by proposing an experiment for their generation.
\end{abstract}

\maketitle

Vortex filaments are one of the most spectacular manifestations of fluids driven out of equilibrium. They are highly-rotating localized zones of fluids, whose extension can be astonishingly large compared to their vortex core \cite{saffmanVortexDynamics1993}. Perhaps one of the most remarkable example in Nature are tornadoes, whose intensity can be devastating \cite{dahlCentrifugalWavesTornadoLike2021}. They also play an important role on safety issues of aircrafts, as they are easily generated by airfoils\cite{spalartAirplaneTraiingVortices1998}. In turbulent flows, they are seen as the most dissipative structures and thus responsible for the violent velocity fluctuations leading to intermittency \cite{sheUniversalScalingLaws1994}. Besides classical fluids, vortex filaments are fundamental hydrodynamic excitations in superfluids, such as $^4$He and Bose-Einstein condensates. In such quantum fluids, vortex filaments are topological defects, with a quantized circulations and establish a bridge between the communities of quantum systems and classical hydrodynamics \cite{feynmanChapterIIApplication1955}.

\begin{figure*}[t]
  \includegraphics[width=1\textwidth]{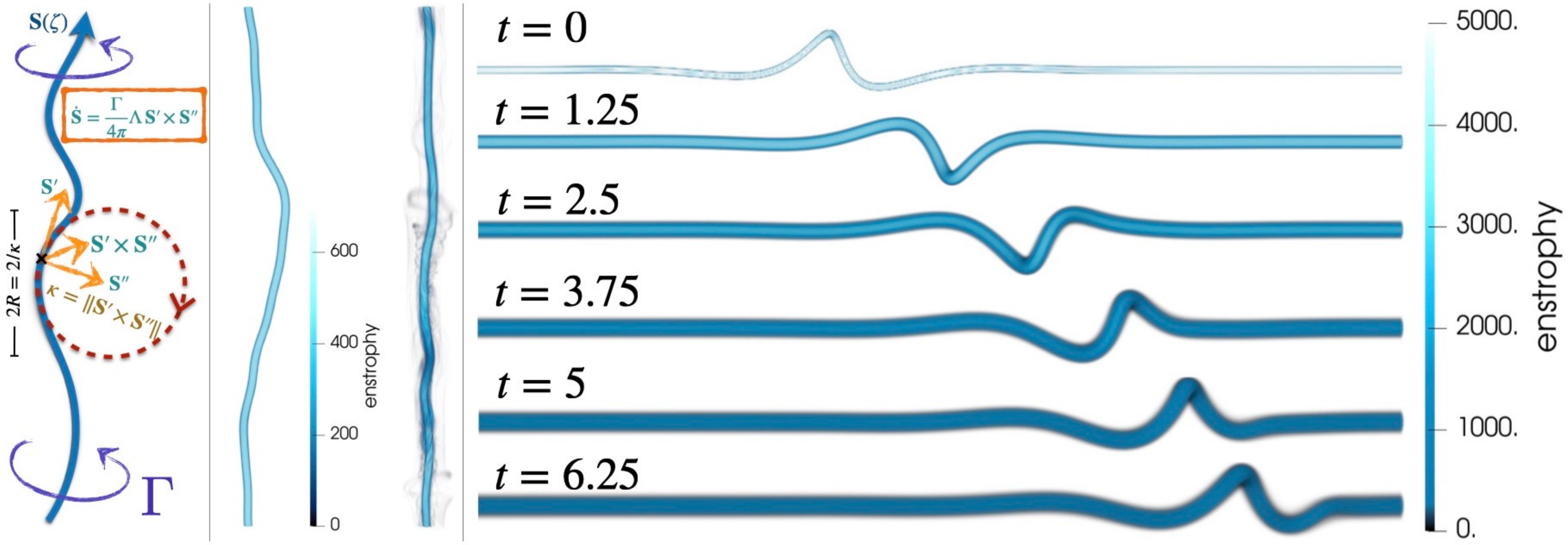}
  \caption{\textbf{(Left)} Sketch explaining the local induced approximation (LIA) model \ref{eq:LIA}.\textbf{(Center)} Kelvin waves propagating in a vortex filament $Re_{\rm v}=5\times 10^{3}$, $L_z/a_0=256$. \textbf{(Right)} A Hasimoto soliton with $A=0.45$, $\lambda=0.65$, $Re_{\rm v}=2.5\times 10^{3}$, $L_z/a_0=512$. Visualizations display three-dimensional rendering of the enstrophy field $|{\bf \omega (x)}|^2$.
    }
    \label{fig:visKW_soliton}
\end{figure*}

The simplest, although crude, description of a vortex filament was proposed by Da Rios in the early 20th century within an approximation known today as the Local Induced Approximation (LIA) \cite{dariosSulMotoLiquido1906,Batchelor_2000}. In this approximation, we consider a vortex filament as a three-dimensional curve ${\bf S}(\zeta)$, with $\zeta$ its arc-length, and having an infinitesimally thin core of size $a_0$. The LIA equation is then written by assuming that each point of the filament roughly moves at the speed and direction of a vortex ring tangent to it and having a radius $R$ equal to the local radius of curvature of the filament. A diagram showing the geometrical derivation of the LIA model is shown in Fig.\ref{fig:visKW_soliton} (left).
Recall that in a perfect fluid, the speed of a ring is given by the general formula $v_{\rm ring}=\frac{\Gamma}{4\pi R}(\log{(R/a_0)}-c)$, where $\Gamma$ is the velocity circulation of the vortex that characterizes its intensity, and $c$ is a order-one constant depending on the specific core model \cite{saffmanVortexDynamics1993}. The LIA equation is then readily written by expressing the radius of curvature in terms of the parametrization and by treating the logarithm in $v_{\rm ring}$ as a constant denoted here $\Lambda$. It reads
\begin{equation}
\partial_t {\bf S}= \frac{\Gamma \Lambda}{4\pi}{\bf S}'\times {\bf S}'', \label{eq:LIA}
\end{equation}
where $'$ denotes derivatives with respect to the arc-length $\zeta$. Note that for a vortex ring of radius $R$, the LIA equation predicts a translation speed $v^{\rm LIA}_{\rm ring}=\Gamma \Lambda /4\pi R$, which misses the logarithmic correction that arrises from non-local contributions of the Biot-Savart law, ignored at this level of approximation. The constant $\Lambda$ is of the order of $\log{\ell/a_0}$, where $\ell$ is a relevant large scale cut-off (e.g. the system size or the inter-vortex distance). This constant appears in the so-called vortex tension $\epsilon =\rho \Gamma^2 \Lambda/4\pi $, with $\rho$ the fluid density. In the LIA framework, it relates the total kinetic energy per unit of mass of the flow $E$ and the total vortex-length as $E=\epsilon \mathcal{L}$ \cite{soninVortexOscillationsHydrodynamics1987}. The actual value of $\Lambda$ is not important in the following as it can be absorbed by renormalizing the time.

Despite the crude approximations, the LIA \eqref{eq:LIA} provides an insightful description of vortex dynamics. The simplest prediction is the propagation of helical excitations along vortex filaments. Using a cartesian parametrization of an almost straight filament ${\bf S}=(X(z),Y(z),z)$, and assuming small deformations, the LIA equations reduces to the (linear) Schrödinger equation $i\partial_t s= \frac{\Gamma \Lambda}{4\pi} \partial_{zz}s$, where $s(z)=X(z)+iY(z)$. It leads to waves with dispersion relation $\omega_k\propto k^2$, where $k$ is the wavevector. Such waves were first introduced in the context of the incompressible Euler equation by Sir~W.~Thomson (also known as Lord Kelvin), and are known today as Kelvin waves (KWs) \cite{thomsonVibrationsColumnarVortex1880}. Considering the full hydrodynamic equations, Lord Kelvin found that these waves have a dispersion relation
\begin{equation}
  \omega_k=\frac{\Gamma}{4\pi}k^2(b - \log{k a_0}), \quad {\rm for \,} a_0 k\ll 1 \label{eq:KWdispRel}
\end{equation}
where $b$ is another order-one core-depending constant. Note that the logarithmic term arises from non-local contributions ignored in the LIA equation, which are somehow encompassed by the constant $\Lambda$. More recent theoretical studies have addressed the role of viscous dissipation and more realistic vortex profiles \cite{ledizesAsymptoticDescriptionVortex2005,fabreKelvinWavesSingular2006}.
Although KWs originate from purely classical hydrodynamics, their importance has been mainly recognized by the superfluid helium and Bose-Enstein condensates community. The reason is twofold. On the one hand, the velocity circulation in supefluids is quantized and vortices are topological defects, having core size that is many orders of magnitude smaller than their typical extension, which makes them perfect vortex filaments from the theoretical point of view. Furthermore, KWs play a crucial role in quantum turbulence. Kelvin waves at different wavelengths interact nonlinearly, making possible the energy transfers towards the smallest scales of superfluid through a wave turbulence cascade\cite{lvovWeakTurbulenceKelvin2010,krstulovicKelvinwaveCascadeDissipation2012,mullerKolmogorovKelvinWave2020}. 
Kelvin waves in superfluids have already been observed experimentally by a few groups \cite{fondaDirectObservationKelvin2014,perettiDirectVisualizationQuantum2023,minowaDirectExcitationKelvin2025}, but measuring their dispersion relation with some precision has remained a challenge. On the other hand, in classical fluids, such as water, KWs can be easily observed, but their dispersion relation has been measured only very recently \cite{BarckickeKW_2026}.

In addition to supporting the propagation of KWs, the LIA equation establishes a bridge between vortex filament dynamics and integrable systems \cite{hasimotoSolitonVortexFilament1972}. Indeed, it is possible to map \eqref{eq:LIA} to the one-dimensional focusing nonlinear Schrödinger equation (1dNLS), a well-known example of integrable system \cite{zakharovExactTheoryTwodimensional1972}. By using the transformation $\psi(\zeta,t)=\kappa(\zeta,t)\exp{-\{i\int \tau(\zeta',t)d \zeta'\}}$, with $\kappa$ and $\tau$ the curvature and torsion of the line respectively, Hasimoto showed that $\psi$ satisfies 
\begin{equation}
i\partial_t \psi=-\frac{\Gamma \Lambda}{4\pi}\left(\partial_{\zeta\zeta}\psi +\frac12 |\psi|^2\psi\right).
\end{equation}

The 1dNLS equations admits the propagation of many types of solitons. The simplest 1dNLS soliton is given by $\psi(\zeta,t)=(2/\lambda) {\rm sech}{[(\zeta - c t)/\lambda]}e^{i \tau_0 s+i\mu t}$, with $\lambda$ the width of the soliton and  $c$ its propagation speed. $\tau_0$ corresponds to the constant torsion of the vortex line and $\mu$ is another constant. This soliton propagates under the 1dNLS equation at constant speed and keeps it shape. Reversing the Hasimoto transformation, leads to the following solitonic solution of the LIA equation
\begin{eqnarray}
\nonumber{\bf S}(\zeta,t)&=&\left(Re[s],Im[s],\zeta - A \tanh{\left( \frac{\zeta- ct}{\lambda}\right)\,}\right)\\
s(\zeta,t)&=& A\,{\rm sech}{\left(\frac{\zeta- ct}{\lambda}\right)}e^{i\left(\frac{2c\pi}{\Gamma \Lambda}(\zeta-ct) +\frac{\Gamma\Lambda}{4\pi \lambda^2}t\right)}\label{eq:Hasimoto_soliton}\\
c&=&\frac{\Gamma \Lambda\tau_0}{2\pi}=\frac{\Gamma \Lambda}{2\pi\lambda}\sqrt{\frac{2\lambda}{A} -1},\quad {\rm with}\, 2\lambda >A.\label{eq:Hasimoto_speed}
\end{eqnarray}
The Hasimoto soliton has been rewritten in terms of  its amplitude $A$, and its propagation speed given as a function of its torsion in \eqref{eq:Hasimoto_speed}. Note that the soliton can be reparametrized in cartesian coordinates if $A<\lambda$. Finally, it is worth mentioning that a single KW in the LIA framework is a constant curvature and torsion helical wave, which corresponds to a plane wave in the mapped 1dNLS. Moreover, other 1dNLS solutions, such as the Ahkmediev breather solitary wave has been numerically studied in different models of superfluids that go beyond the LIA approximation \cite{salmanBreathersQuantizedSuperfluid2013}.

A natural question arises. Is it possible to observe such type of vortex excitations in classical viscous vortex filaments, such as those in water or air? More than forty years ago, Hopfinger and Browand \cite{hopfingerVortexSolitaryWaves1982} observed a travelling disturbance over a vortex filament in a rotating turbulent water tank experiment. Such a disturbance was compatible with a Hasimoto soliton, but we lack today of a clear understanding of the mechanisms that could generate it and how to trigger it in a controlled and reproducible manner.
The approximations leading to the LIA equation are certainly very strong, but one may ask under which conditions its predictions remain, at least phenomenologically, valid. Whereas the assumption of infinitesimally small vortex core is reasonable, provided that the scales of observation are much larger than $a_0$, the assumption of locality might be strong. Indeed, within the LIA description, two approaching segments of a vortex do not see each other and can cross, which is a non-physical situation, as for real vortices this configuration would lead to a vortex reconnection. Finally, in real viscous flows, the vortex dynamics will not only result from the interplay between dispersion and nonlinearity, but also from viscous dissipation. 

In this work, we address the aforementioned questions by studying the propagation of Kelvin waves and solitons in classical vortex filaments described by the three-dimensional incompressible Navier-Stokes (NS) equations
\begin{eqnarray}
    \frac{\partial\vvec}{\partial t}+\vvec \cdot\nabla \vvec &=&-\nabla p +\nu \nabla^2\vvec\label{eq:NS}\\
    \nabla\cdot\vvec&=&0,\label{eq:NSdiv}
\end{eqnarray}
where $\vvec$ is the velocity field, $p$ the pressure, $\nu$ the kinematic viscosity of a fluid having a unit density.
We consider a vortex filament, which initial vorticity field ${\bf \omega}=\nabla\times\vvec$ is given by the line integral
\begin{equation}
    {\bf \omega}({\bf x}) = \Gamma \oint \delta_{a_0}({\bf x - S}){\rm d}{\bf S},\label{eq:vorticity}
\end{equation}
where $\delta_{a_0}()$ is a regularization of the Dirac-$\delta$ defining the vortex core model. 

We are naturally interested in the limit where the vortex Reynolds number $Re_{\rm v}=\Gamma/\nu\gg1$, and $a_0$ is much smaller than any other length scale in the system. Given a vortex parametrization ${\bf S}$, \eqref{eq:vorticity} provides the associated velocity field by inverting the curl operator that we then use as initial condition for the NS Eq.(\ref{eq:NS}-\ref{eq:NSdiv}). We integrate the NS equation using the pseudo-spectral code GHOST, with a second-order Runge-Kutta time stepping method and we use resolutions of $512^2\times1024$ grid points. We consider a domain of size $L\times L\times L_z$, and filaments oriented along the $z-$axis. To ensure periodicity, we set four vortices of alternated signs. In all simulations, we use a solid-body rotation model for the core.
Finally, we track the vortex position on-the-fly by using a weighted average.
See \ref{App:methods} and \ref{App:supplemental} for more details.

\begin{figure}[h!]
  \centering
  \includegraphics[width=.95\columnwidth]{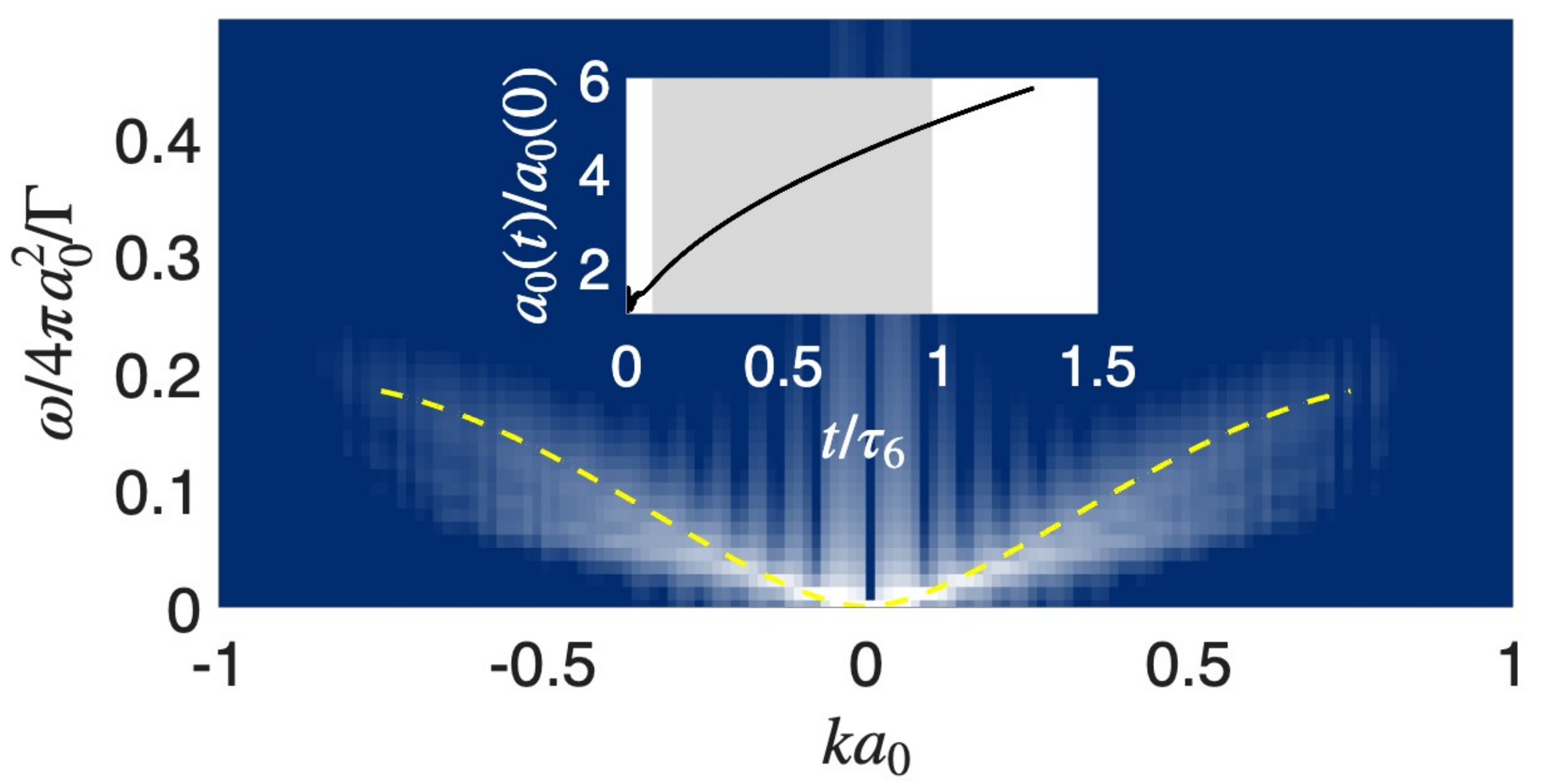}
  \caption[]{\textbf{(Left)} Kelvin waves propagating in a vortex filament $Re_{\rm v}=1.7\times 10^{3}$, $L_z/a_0=512$. Kelvins wave have a global amplitude of $C_0=0.07$ (see text). The white dashed line is the KW asymptotic prediction \eqref{eq:KWdispRel} and the green dashed line is the full dispersion relation given in the SI. The inset displays the evolution of the core size, where $\tau_6=2\pi/\omega_{k_6}$, with $k_6=6\times2\pi/L_z$. The dispersion relation has been calculated in the grey area.
    }
    \label{fig:KWdisp_relation}
\end{figure}

We start by studying the propagation of KWs in an almost straight vortex, with perturbations at all scales given by $s(z)=C_0\sum_k k^{-1} e^{i(k\,z+\phi_k)}$, where $\phi_k$ are random phases, and $C_0$ fixes the amplitude. The evolution of large amplitude Kelvin waves is displayed in Fig.~\ref{fig:visKW_soliton} (center) as an example. 
To measure the KW dispersion relation, we set small amplitude KWs with wavelengths spanning almost all available scales. Once the vortex filament is tracked, we Fourier transform in time and space $s(z,t)$ and compute its magnitude. We also take advantage of the four vortices present in the computational box to reduce statistical noise through average. In Fig.\ref{fig:KWdisp_relation} we present the dispersion relation in excellent agreement with the Lord Kelvin's theoretical prediction for an inviscid solid-body-rotation core vortex filament.
The white dashed line corresponds to the asymptotic Kelvin wave dispersion relation \eqref{eq:KWdispRel} with $b=\log(2)-\gamma_{\rm E} + 1/4 $, where $\gamma_{\rm E}=0.5772\ldots$ is the Euler–Mascheroni constant. The green dashed line shows the full theoretical prediction given by an implicit formula involving special functions (see \cite{KrstulovicHDR} and \ref{App:supplemental}). We use the initial value of the vortex core in the analytical formulas. During the temporal evolution, as displayed in the inset, the vortex core increases its size due to viscous effects, explaining the blurring of the curves.
We have thus demonstrated now that our setting allows for studying Kelvin waves on the filament without being much affected by viscous effects.

\begin{figure}[h!]
  \centering
  \includegraphics[width=1\columnwidth]{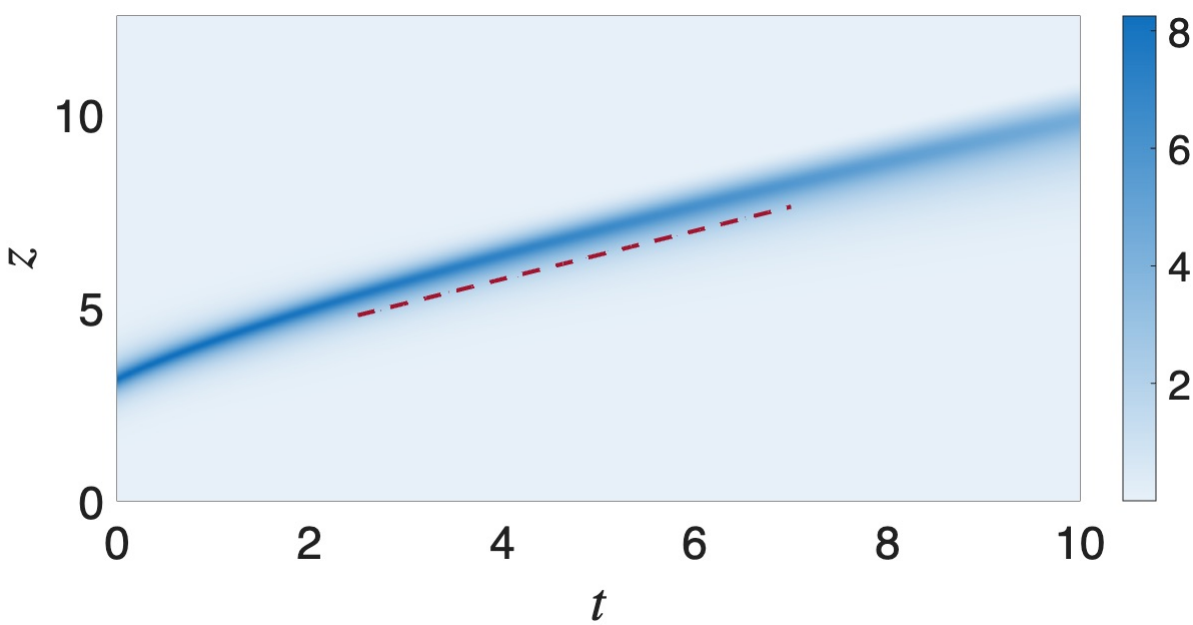}
  \includegraphics[width=1\columnwidth]{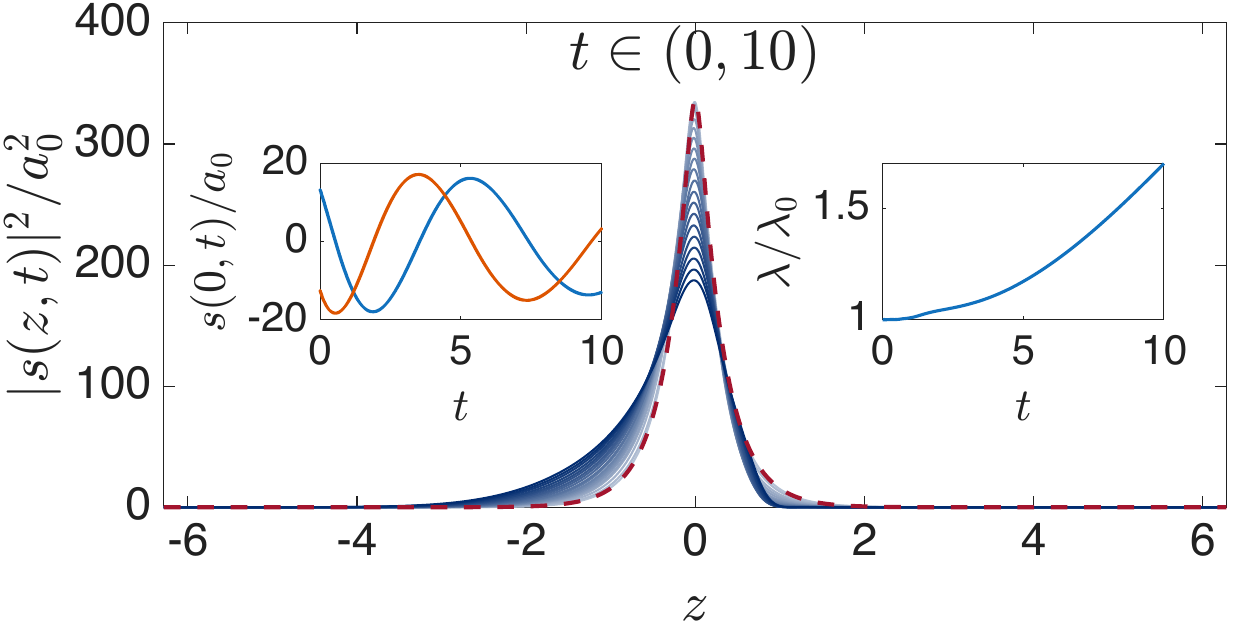}
  \caption[]{\textbf{(Top)} Hasimoto soliton propagating in a vortex filament with $A=0.45$, $\lambda=0.65$, $Re_{\rm v}=2.5\times 10^{3}$, $L_z/a_0=512$ (same of Fig.\ref{fig:visKW_soliton}). \textbf{(Bottom)} Soliton profile centered at $z=0$. The left inset displays the real and imaginary parts of the soliton amplitude at its maximum ($z=0$), showing a coherent oscillation. The right inset shows the width of the soliton as a function of time.  
    }
    \label{fig:SolitonPropagation}
\end{figure}

We now turn to study the propagation of solitons in Navier-Stokes. We set a vortex with a Hasimoto soliton, as displayed in Fig.\ref{fig:visKW_soliton} (right).
While times evolves, the soliton travels oscillating and keeping its shape. In Fig.\ref{fig:SolitonPropagation}.a we present a $z-t$ diagram of the soliton amplitude.
We observe that after a short transient, the soliton travels almost at constant speed, as shown by the dashed red line obtained by fitting the soliton speed. Using this fit, the theoretical expression (\ref{eq:Hasimoto_soliton}), and the measured values of $A$ and $\lambda$ at those times, we fix the phenomenological LIA constant to $\Lambda_{\rm fit}=1.857$, value that we will use later.
Figure \ref{fig:SolitonPropagation}.b displays its amplitude as a function of time, showing that its amplitude decays slowly while increasing its width. The profiles have been centered at the origin. The red dashed line is the theoretical Hasimoto soliton used to set the initial condition. The insets show the temporal evolution of the real and imaginary parts of $s(0,t)$ (left inset) and the width of the soliton. Note that despite viscous effects, the soliton has travelled more than ten times it original width. We expect that by increasing the $Re_{\rm v}$ number, the soliton should preserve its shape better.

\begin{figure}[h!]
  \centering
  \includegraphics[width=1\columnwidth]{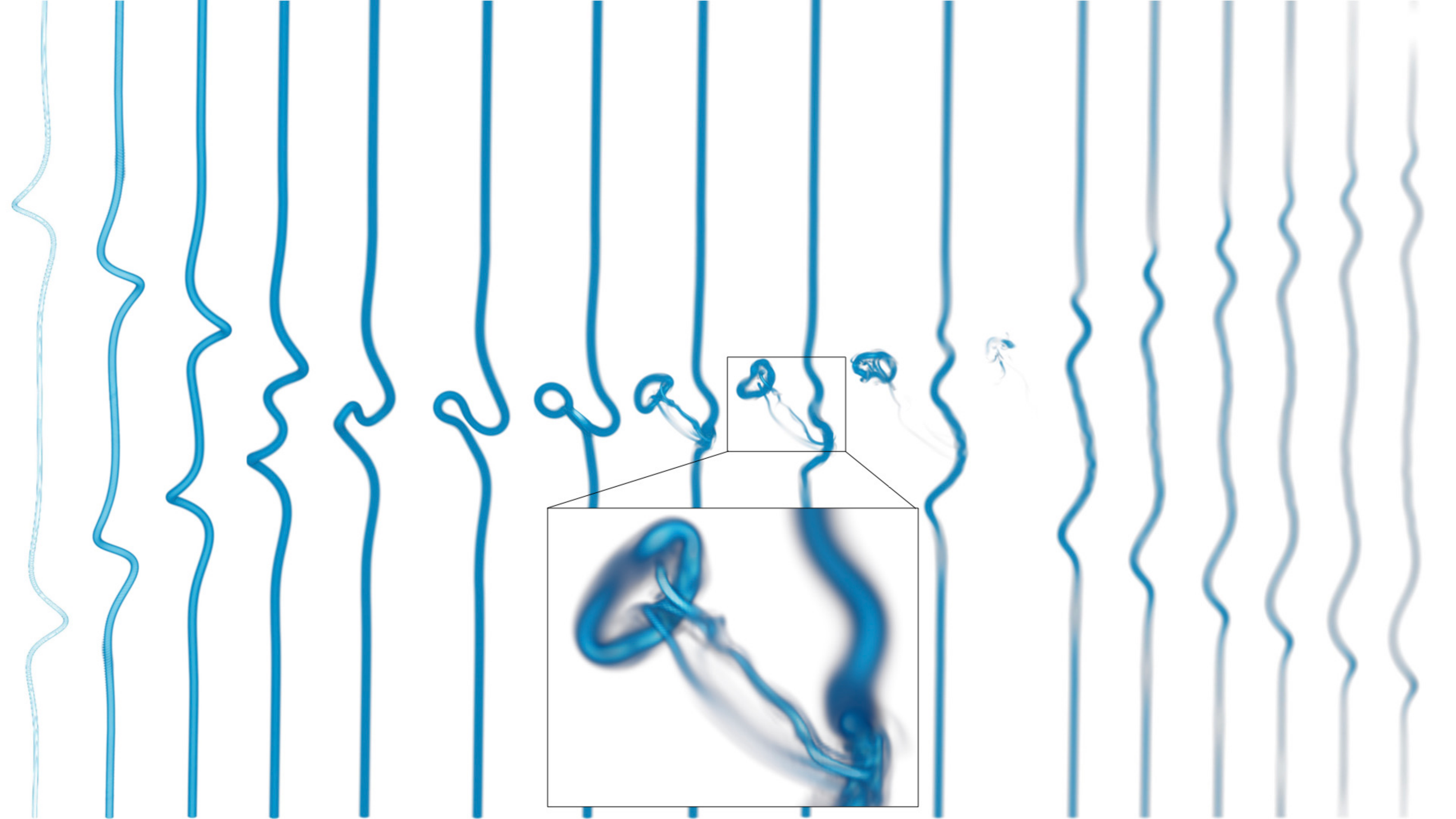}
  \includegraphics[width=1\columnwidth]{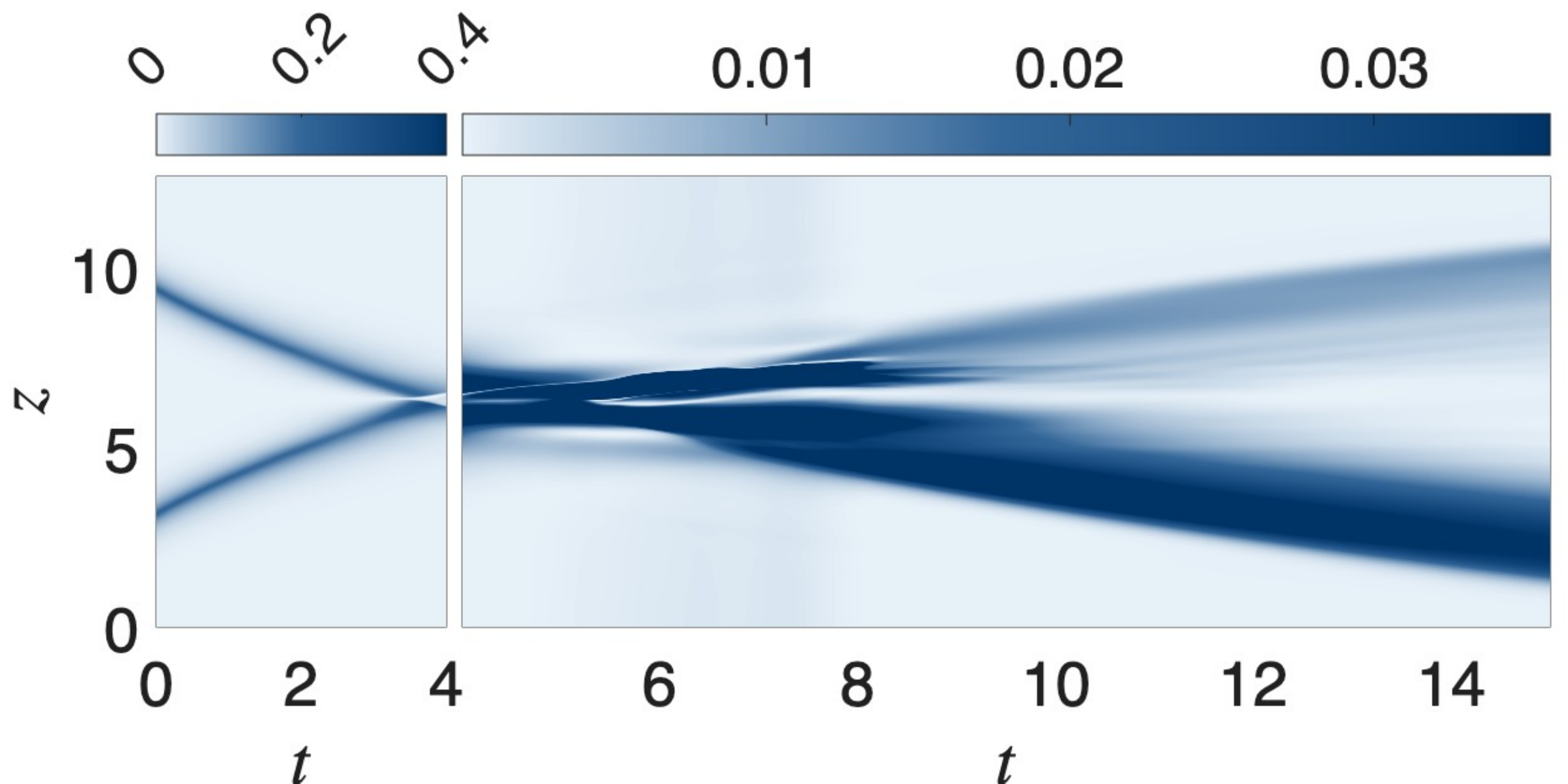}
  \caption[]{ Collision of two Hasimoto solitons propagating in a vortex filament with $A=0.45$, $\lambda=0.65$, $Re_{\rm v}=2.5\times 10^{3}$, $L_z/a_0=512$. \textbf{(Top)} Three-dimensional rendering of the enstrophy field $|{\bf \omega (x)}|^2$ showing the two colliding vortex solitons with a close up into the emission of a vortex ring. \textbf{(Bottom)} $z-t$ diagram of the solitons showing their collision and the two surviving solitons.}
    \label{fig:soliton_colission}
\end{figure}

One of the most striking characteristic of 1dNLS solitons is that when they meet, they interact elastically, i.e. two solitons traveling towards each other, after a short time of interaction, keep traveling keeping the same properties. In a classical fluid, we can not expect a perfect elastic collision as the system is not conservative. Furthermore, the existence of the solitons reposes on the LIA description, which ignores long-range vortex interactions with other sections of the filaments. We investigate this problem, by setting two conjugate solitons initially well separated, as shown in Fig.\ref{fig:soliton_colission}.a.
During the first part of the evolution, as expected, the two solitons travels towards each other until they collide. During the collision, the filaments twist and two segments get very close. As these two portions of the filaments have different signs, a vortex reconnection is triggered, and a vortex ring is then ejected from the filament. Note that vortex bridges join the filament to the ring. Those very thin structures are quickly dissipated by viscosity. After some time, two smaller surviving solitons travel away from each other.  Fig.\ref{fig:soliton_colission}.b displays the temporal evolution of the solitons in a $z-t$ diagram, where the evolution is clearer. Note that after the collision the colormap indicates that the solitons are roughly five times smaller. Also note that our tracking method is not reliable during the collision as the secondary structures modify the weighted average \eqref{eq:weightedAve}. A LIA simulation of this configuration is given for sake of completeness in the SI. As expected, within the LIA model, the two solitons traverse each other without affecting their shape.

\begin{figure}[h!]
  \centering
  \includegraphics[width=1\columnwidth]{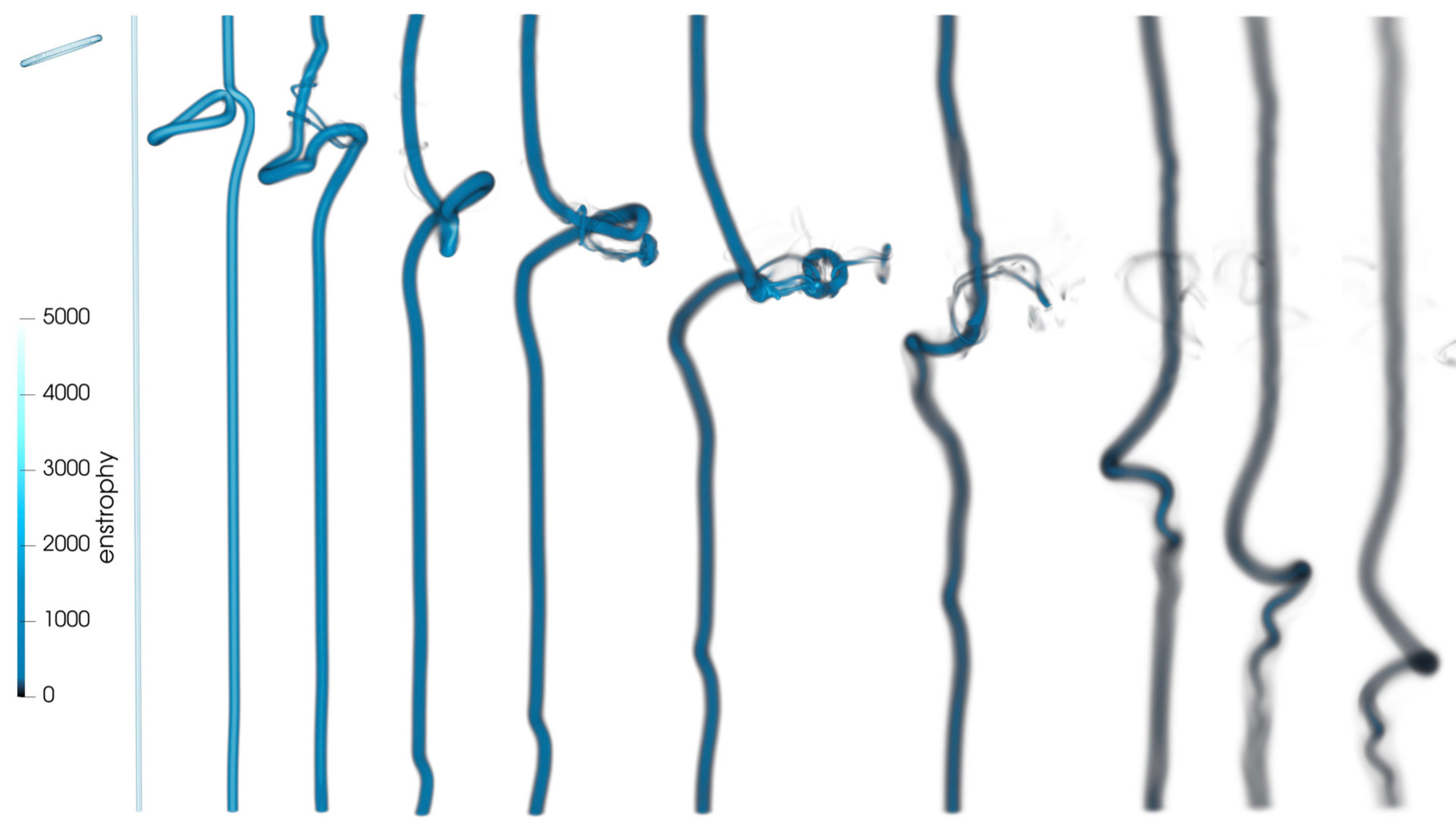}
  \includegraphics[width=1\columnwidth]{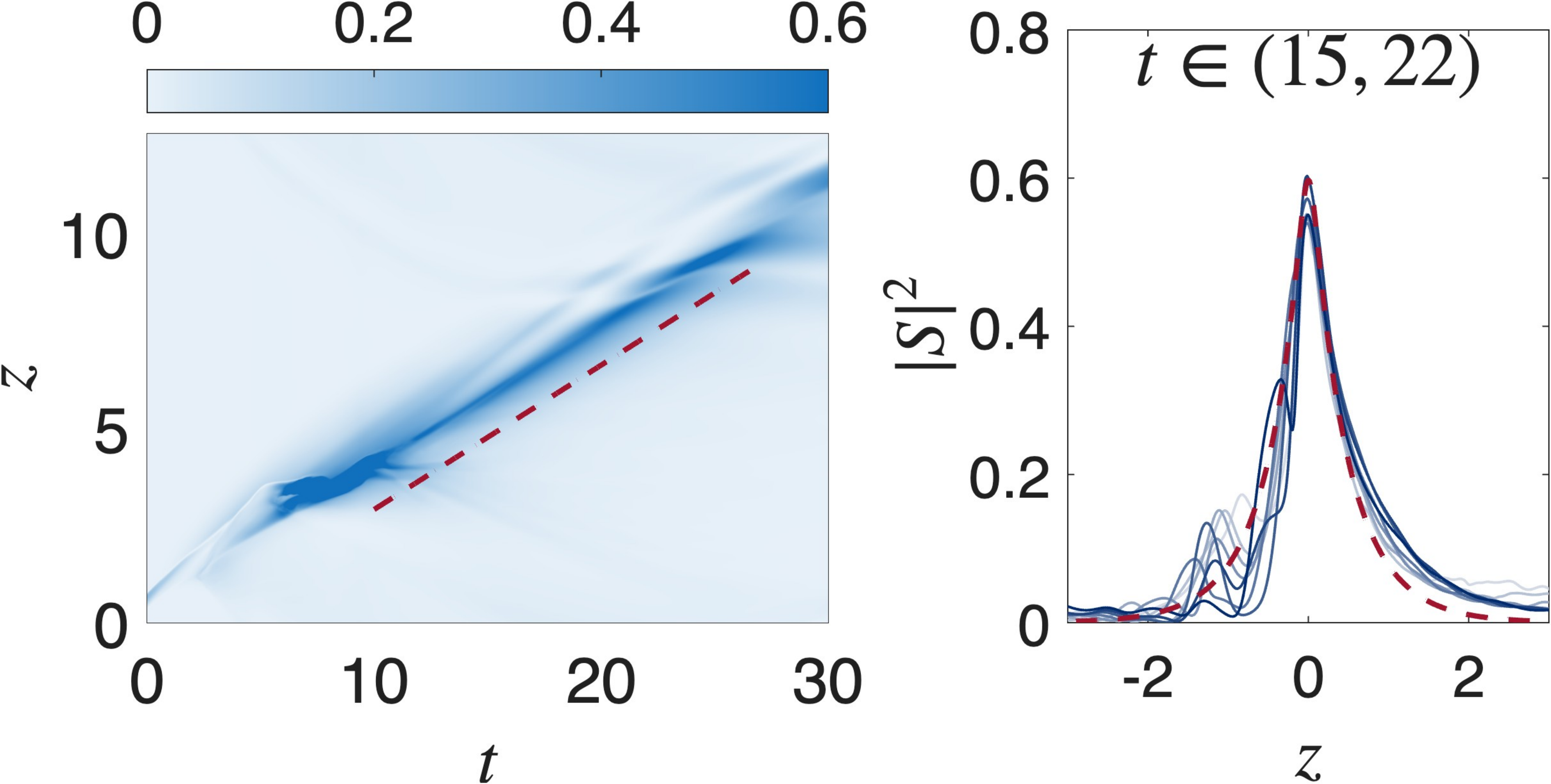}
  \caption[]{Generation of a soliton by the collision of a vortex ring of radius $R=0.6$ (visible at the top left corner), both the ring and the filament have a $Re_{\rm v}=2.5\times 10^{3}$, $L_z/a_0=512$.\textbf{(Top)} Three-dimensional rendering of the enstrophy field $|{\bf \omega (x)}|^2$.  \textbf{(Bottom left)} $z-t$ diagram of the soliton amplitude. Collision takes place around $t=8$. \textbf{(Bottom right)} Soliton profile centered at $z=0$ for different times. 
    }
    \label{fig:soliton_generation}
\end{figure}

Finally, we address a practical question: would it be possible to generate such a soliton in an experiment? A vortex soliton is a nonlinear structure traveling along the vortex axis, and thus it carries momentum in that direction. We expect that if a vortex filament receives a sudden transfer of momentum in the direction of its orientation, a soliton could be triggered on the filament to absorb the excess of momentum. In order to test this idea, we throw a small vortex ring towards the filament. The initial condition of our numerical experiment is displayed in Fig.\ref{fig:soliton_generation}.a. 
The ring travels towards the filament, getting closer and reconnecting. After the reconnection, thin vortex structures are created which are quickly dissipated, but the filament has clearly acquired some vertical momentum. As a result, a vortex soliton is generated, carrying the momentum transferred by the ring. Figure \ref{fig:soliton_generation} (bottom left) displays the $z-t$ diagram clearly showing the propagation of the soliton. Finally, Fig.\ref{fig:soliton_generation} (bottom right) shows the shape of the soliton, which is well preserved during its evolution.

To provide quantitative estimates of the generated soliton, we recall that the momentum of a vortex ring of radius $R$ along its propagation axis is $P_{\rm ring}=\Gamma_{\rm ring} \pi R^2$, and its energy in the LIA framework is $E_{\rm ring}=2\pi R\,\Gamma^2_{\rm ring}\,\rho\Lambda/4\pi$. Similarly, the soliton carries a momentum (see Methods) along it axis equal to $P_{\rm sol}= \Gamma A^2 \sqrt{\frac{2\lambda}{A} -1}$ and it has an extra length (with respect to a straight vortex) leading to an extra energy $\Delta E_{\rm sol}\approx 2A\,\Gamma^2\,\rho\Lambda/4\pi$. By imposing that an order-one fraction of the vortex ring momentum and energy is transferred to the soliton, i.e. $\Delta E_{\rm sol}=\gamma_1 E_{\rm ring}$ and $P_{\rm sol}=\gamma_2 P_{\rm ring}\cos{\alpha}$, where $\alpha$ is the angle of orientation of the ring with respect to the straight vortex, and $0<\gamma_1,\gamma_2\le1$ quantify the losses. It follows that
\begin{equation}
  A=\gamma_1\frac{\Gamma_{\rm ring}^2}{\Gamma^2}\pi R,\quad \lambda = \frac{A}{2}\left(1 + \frac{\gamma_2^2}{\gamma_1^4}\frac{ \Gamma^6}{\Gamma_{\rm ring}^6}\frac{\cos^2{\alpha}}{\pi^2}\right).\label{eq:A_lambda_vs_Ring}
\end{equation}
Note that the condition $A\le 2\lambda$ is always fulfilled.

Figure \ref{fig:soliton_generation} (bottom right) also shows as a dashed red line the Hasimoto soliton using the above formulas with the ring parameters and fitted with $\gamma_1=0.41$ and $\gamma_2=0.32$. Remarkably, the dashed red line on Fig.~\ref{fig:soliton_generation} (bottom left) shows the theoretical soliton speed \eqref{eq:Hasimoto_speed} using $\Lambda=\Lambda_{\rm fit}$, measured from the the single soliton in Fig.\ref{fig:SolitonPropagation}. The agreement is excellent despite the simplicity of the estimate. Note that in general one could expect that $\gamma_1,\gamma_2$ might be functions of $R/a_0$, $\Gamma/\nu$, and the ratio of the ring and the straight vortex circulation. Determining such functional dependence is out of the scope of this work, but the relations provided in \ref{eq:A_lambda_vs_Ring} remain a simple and useful estimate.

In this work, we have shown that vortex filaments in classical viscous fluids exhibit several stable excitations, namely Kelvin waves and solitons. Despite the strong assumptions of the LIA model \eqref{eq:LIA}, its simple geometric foundations make it robust and allow qualitative prediction of such vortex excitations. The LIA model has attracted the interest of physicists and mathematicians for more than a century. It brings together the physics and mathematics of dispersive systems, fluid mechanics and a powerful connection with integrable systems. We have also proposed, and demonstrated numerically, a possible method for generating solitons in laboratory experiments.

This work opens many research avenues. Following a traditional fluid-mechanics approach, one can attempt to directly prove the existence of vortex solitons in the Navier-Stokes equation, ask under which conditions they are stable, and how they interact with other Kelvin wave modes not included in the LIA framework. From a practical perspective, one still needs to characterize and better understand how to trigger solitons on vortex filaments. In this work, we have shown that vortex reconnections can generate solitons. By this mechanism, not only a net transfer of energy and vertical momentum occur, but the vortex also acquires a non-zero helicity. It is well known that helicity plays an important role as a proxy for tornado forecasts. One might then ask whether solitons naturally arise in highly helical flows.

Finally, by showing that Kelvin waves can propagate without much damping in classical vortices at moderate Reynolds numbers, we open the possibility of applying the theoretical predictions of wave turbulence to the Kelvin wave cascade, which predicts an energy transfer towards small scales. Indeed, to date, the Kelvin wave has been mainly studied theoretically and numerically in the context of superfluids, as quantum vortices are topological defects, which makes them extremely stable. Their study in cryogenic experiments, however, remains an enormous challenge. This work, together with the very recent experimental measurement of the Kelvin wave dispersion relation in water, not only reveals a new experimental setting to test wave turbulence predictions, but also unveils the relevance of studying a nonlinear wave mechanism for energy transfer towards small scales usually neglected in classical vortex filaments.

\appendix

\section{Methods \label{App:methods}}
\subsection{Navier-Stokes vortex initial condition}
In order to generate the initial condition for the NS \eqref{eq:NS}, we numerically evaluate the Fourier transform of ${\bf \omega}({\bf x})$, \eqref{eq:vorticity}, as
\begin{equation}
\hat{\bf \omega}({\bf k}) =\Gamma\oint \hat\delta_{a_0}(|{\bf k}|)e^{i{\bf k}\cdot {\bf S}} {\rm d}{\bf S}.
\end{equation}
Then, the velocity field is computed in Fourier space as $\hat{\bf v}({\bf k})=i{\bf k}\times \hat{\bf \omega}({\bf k})/|{\bf k}|^2$. For the core model, we use a solid body rotation that explicitelly reads in Fourier space $\hat\delta_{a_0}(k)=3 (\sin{(a_0 k)} - a_0 k \cos{(a_0 k)} )/(a_0 k)^3$. We use a box of size $L=2\pi$, and $L_z=4\pi$.

\subsection{Vortex length and momentum}
The momentum of a vortex line is readily computed using the well-known expression ${\bf P}=\frac{\Gamma}{2}\oint {\bf S} \times \mathrm{d}{\bf S}$ \cite{soninVortexOscillationsHydrodynamics1987,Pismen:1999aa}, and the parametrization of a ring and the one of the soliton. The total extra length of a soliton with $\lambda\ll L_z$ is simply given by
\begin{equation*}
\Delta L_{\rm sol}=L_z-(S_3(L_z/2)-S_3(-L_z/2))=2A\tan{\frac{L_z}{2\lambda}}\approx 2A.
\end{equation*}

\subsection{Soliton in the periodic box}
In order to properly set a soliton in the middle of the periodic box of height $L_z$, we shift the $z$-component in \eqref{eq:Hasimoto_soliton} as  
\begin{equation}
{\bf S}_{\rm box}(\zeta) = (S_1,S_2,S_3+L_z/2), \quad \zeta\in(-\zeta_{\rm lim},\zeta_{\rm lim}),
\end{equation}
where $\zeta_{\rm lim}=A \tanh{(L_z/\lambda)} + L_z/2$ is the total arclength of the soliton from the bottom of the box to $L_z/2$. With this definition, ${\bf S}_{\rm box}(-\zeta_{\rm lim})\approx (0,0,0)$ and ${\bf S}_{\rm box}(\zeta_{\rm lim})\approx (0,0,L_z)$, provided that $\lambda\ll L_z$. We then shift the vortex in the $xy$-plane to $(L/4,L/4)$, and add three other alternating-sign straight vortices in the symmetric positions in order to ensure periodicity. For the two solitons initial conditions, we apply the same method in the bottom half of the box and glue another conjugate soliton in the upper half.

\subsection{Vortex measurements}
The vortex positions are measured by the vorticity-weighted average
\begin{equation}
 s(z,t)=\int_\mathcal{D} (x +i y)|\omega(x,y,z)|^2\mathrm{d}x\mathrm{d}y\Big/\int_\mathcal{D}|\omega(x,y,z)|^2\mathrm{d}x\mathrm{d}y\label{eq:weightedAve}
\end{equation}
where $\mathcal{D}$ is a subdomain containing the vortex.

\section{Supplemental materials \label{App:supplemental}}

\subsection{Kelvin wave dispersion relation}

In its original work, Lord Kelvin derived the dispersion relation of waves propagating along a straight vortex \cite{thomsonVibrationsColumnarVortex1880}. More precisely, he considered the following solution of the incompressible three-dimensional Euler equation
\begin{eqnarray}
{\bf v}(r,\theta,z)&=&\frac{\alpha(r)}{r}\hat{\theta}\\
 p(r,\theta,z)&=&p_0(r)=\rho_0\int_{a_0}^r\frac{\alpha(s)^2}{s^3}\mathrm{d}s. \label{KWEq:defVortex}
\end{eqnarray}
where ${\bf v}(r,z,\theta)$ and $p(r,z,\theta)$ are the velocity and pressure fields expressed in cylindrical coordinates respectively. The function $\alpha(r)$ sets the total circulation and the vortex profile, including its core size denoted as $a_0$. By perturbing this solution and choosing some vortex profile yield the vortex wave dispersion relation. See reference \cite{KrstulovicHDR} for a derivation using a modern notation.
In particular, by choosing the solid-body rotation profile with circulation $\Gamma$
\begin{equation}
\alpha(r)= \frac{\Gamma}{2\pi}\begin{cases}
		 \frac{r^2}{a_0^2} &\text{if }r \le a_0 \\
                 1   &\text{if }r \ge a_0.
                \end{cases}
\end{equation}
leads, after some algebra, to an implicit relation for the wave frequency $\omega$, the wave vector along the $z$-direction $k$, and the azimuthal wave number $n$. It reads
\begin{widetext}
\begin{equation}\label{KWEq:KWdisprelSolid}
\begin{cases}
\frac{2n}{2+n-\tilde{\omega}}+\frac{|k| K_{n-1}(|k|a_0)}{K_{n}(|k|a_0)}+\frac{2n\, _0F_1\left[n;\frac{1}{4}a_0^2k^2\left(1-\frac{4}{(n-\tilde{\omega})^2}\right)\right]}{\left(1-\frac{4}{(n-\tilde{\omega})^2}\right)\, _0F_1\left[n+1;\frac{1}{4}a_0^2k^2\left(1-\frac{4}{(n-\tilde{\omega})^2}\right)\right]}=0 &\text{if }n\neq 0\\
\\
\frac{|k| K_{1}(|k|a_0)}{K_{0}(|k|a_0)}+\frac{a_0 ^2k^2\,_0F_1\left[2;\frac{1}{4}a_0^2k^2\left(1-\frac{4}{\tilde{\omega}^2}\right)\right]}{2 I_0\left( a_0k \sqrt{1-\frac{4}{\tilde{\omega}^2}}  \right)}=0 &\text{if }n= 0,
\end{cases}
\end{equation}
\end{widetext}
where $_0F_1[\nu;z]$ is the confluent hypergeometric function and $\tilde{\omega}=\omega/(\Gamma/2\pi a_0^2)$. In the literature, Kelvin waves typically refer to the case $n=1$ that in the limit of $a_0k\ll1$ simplifies to
\begin{equation}
\omega(k)=-\frac{\Gamma}{8\pi}k^2\left (\log{\frac{1}{a_0|k|}}+b\right),\label{KWEq:KWdisprelSolidAsymp}
\end{equation}
 with $b=-\gamma_{\rm E}+\log{2}+\frac{1}{4}=0.366\ldots$ and $\gamma_{\rm E}=0.577216\ldots$ the Euler constant, which corresponds to Eq.\eqref{eq:KWdispRel} of the main text.

\begin{figure*}[t!]
\centering
\includegraphics[width=.99\textwidth]{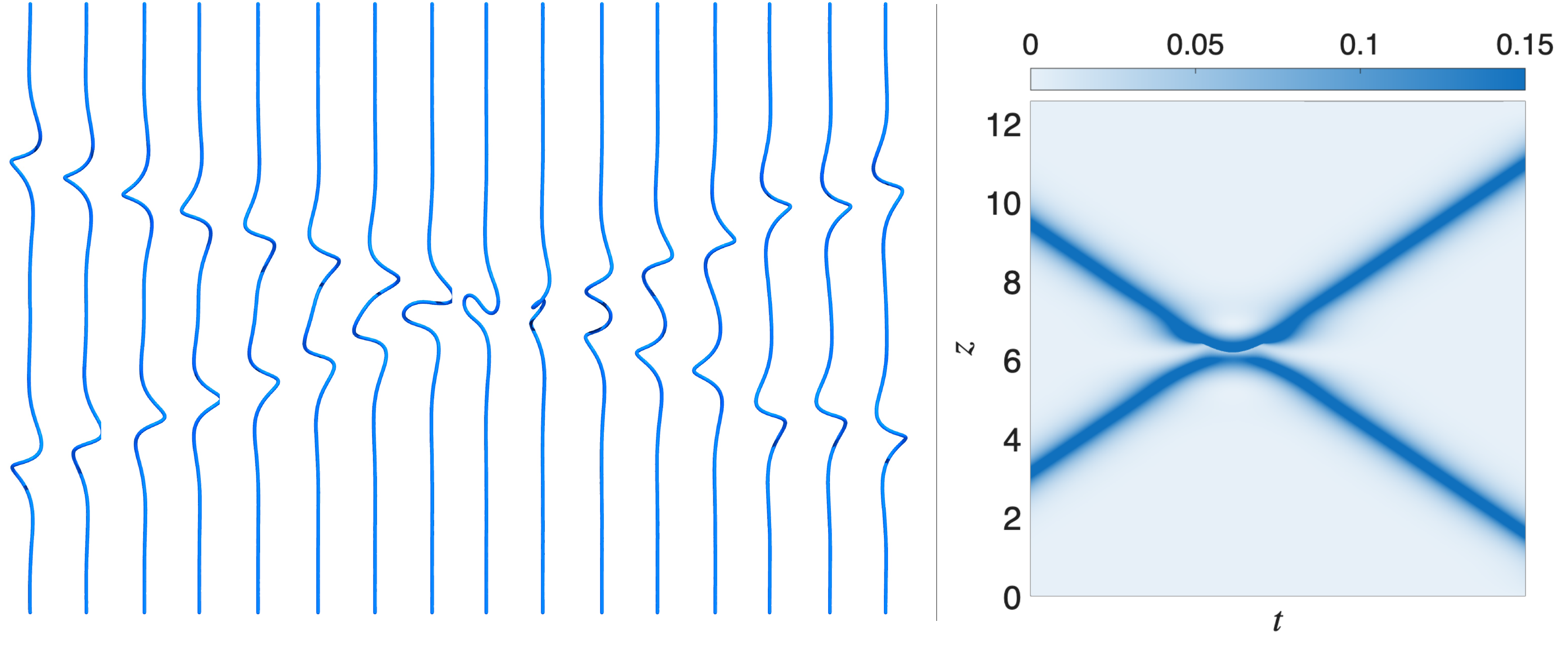}
\caption{\textbf{(Left)} Visualizations of two colliding solitons under LIA dynamics. \textbf{(Right)} $z-t$ diagram of the soliton amplitude.\label{fig:LIAtwoSols}}
\end{figure*}

\subsection{Two Hasimoto solitons within the local induced approximation}

For completeness, we present here the temporal evolution of two colliding Hasimoto solitons evolved under the LIA dynamics. We use the same soliton as in figures 1, 3 and 4, and integrate Eq.[1] numerically. It dynamics is shown in Fig.~\ref{fig:LIAtwoSols} left. The panel on the right displays a $z-t$ diagram showing the amplitude of the vortex perturbation. As expected for two 1dNLS solitons, they keep their shape and size after the collision, as in an elastic ``collision''.

\begin{acknowledgments}
This work was funded by the Simons Foundation Collaboration grant Wave Turbulence (Award ID 651471) and the Agence Nationale de la Recherche through the project QuantumVIW ANR-23-CE30-0024-02. Computations were also carried out at the Mésocentre SIGAMM hosted at the Observatoire de la Côte d'Azur and Azzurra of the OPAL infrastructure from Université Côte d’Azur, supported by the French government, through the UCAJEDI Investments in the Future project managed by the National Research Agency (ANR) under Reference No. ANR-15-IDEX-01.
\end{acknowledgments}

\bibliographystyle{apsrev4-2}
\bibliography{refs_KWsSolitons.bib}

@phdthesis{KrstulovicHDR,
author={Giorgio Krstulovic},
title={A theoretical description of vortex dynamics in superfluids. Kelvin waves, reconnections and particle-vortex interaction},
school={Universite C\^ote d’Azur},
type = {Habilitation \`a diriger des recherches},
month={October},
URL={https://gkrstulovic.gitlab.io/thesisms/hdr-krstulovic/},
year={2020}
}

@article{hopfingerVortexSolitaryWaves1982,
	title = {Vortex solitary waves in a rotating, turbulent flow},
	volume = {295},
	copyright = {http://www.springer.com/tdm},
	issn = {0028-0836, 1476-4687},
	url = {https://www.nature.com/articles/295393a0},
	doi = {10.1038/295393a0},
	language = {en},
	number = {5848},
	urldate = {2026-01-03},
	journal = {Nature},
	author = {Hopfinger, E. J. and Browand, F. K.},
	month = feb,
	year = {1982},
	pages = {393--395},
}

@article{BarckickeKW_2026,
	title = {Kelvin wave propagation along vortex cores},
	issn = {1745-2473, 1745-2481},
	url = {https://www.nature.com/articles/s41567-026-03175-w},
	doi = {10.1038/s41567-026-03175-w},
	language = {en},
	urldate = {2026-02-25},
	journal = {Nature Physics},
	author = {Barckicke, Jason and Falcon, Eric and Gissinger, Christophe},
	month = feb,
	year = {2026},
}

@book{Pismen:1999aa,
	Author = {Pismen, Len M and Pismen, Len M},
	Date = {1999},
	Publisher = {Oxford University Press},
	Title = {Vortices in nonlinear fields: from liquid crystals to superfluids, from non-equilibrium patterns to cosmic strings},
	Volume = {100},
	Year = {1999}}

@article{zakharovExactTheoryTwodimensional1972,
	title = {Exact {Theory} of {Two}-dimensional {Self}-focusing and {One}-dimensional {Self}-modulation of {Waves} in {Nonlinear} {Media}},
	volume = {34},
	issn = {1063-7761},
	url = {https://ui.adsabs.harvard.edu/abs/1972JETP...34...62Z},
	urldate = {2025-11-06},
	journal = {Soviet Journal of Experimental and Theoretical Physics},
	author = {Zakharov, V. E. and Shabat, A. B.},
	month = jan,
	year = {1972},
	note = {Publisher: Springer
ADS Bibcode: 1972JETP...34...62Z},
	pages = {62},
}

@article{salmanBreathersQuantizedSuperfluid2013,
	title = {Breathers on {Quantized} {Superfluid} {Vortices}},
	volume = {111},
	issn = {0031-9007, 1079-7114},
	url = {https://link.aps.org/doi/10.1103/PhysRevLett.111.165301},
	doi = {10.1103/PhysRevLett.111.165301},
	language = {en},
	number = {16},
	urldate = {2020-04-06},
	journal = {Physical Review Letters},
	author = {Salman, Hayder},
	month = oct,
	year = {2013},
	pages = {165301},
}

@book{Batchelor_2000, 
place={Cambridge}, 
series={Cambridge Mathematical Library}, 
title={An Introduction to Fluid Dynamics}, 
publisher={Cambridge University Press}, 
author={Batchelor, G. K.}, 
year={2000}, 
collection={Cambridge Mathematical Library}}

@article{perettiDirectVisualizationQuantum2023,
	title = {Direct visualization of the quantum vortex lattice structure , oscillations, and destabilization in rotating $^{\textrm{4}}$ {He}},
	volume = {9},
	issn = {2375-2548},
	url = {https://www.science.org/doi/10.1126/sciadv.adh2899},
	doi = {10.1126/sciadv.adh2899},
	number = {30},
	journal = {Science Advances},
	author = {Peretti, Charles and Vessaire, Jérémy and Durozoy,Emeric and Gibert, Mathieu},
	month = jul,
	year = {2023},
	pages = {eadh2899},
}

@article{thomsonVibrationsColumnarVortex1880,
	title = {Vibrations of a columnar vortex},
	volume = {10},
	url = {https://www.tandfonline.com/doi/full/10.1080/14786448008626912},
	doi = {10.1080/14786448008626912},
	number = {61},
	urldate = {2020-02-07},
	journal = {The London, Edinburgh, and Dublin Philosophical Magazine and Journal of Science},
	author = {Thomson, William},
	month = sep,
	year = {1880},
	pages = {155--168},
}

@article{soninVortexOscillationsHydrodynamics1987,
	title = {Vortex oscillations and hydrodynamics of rotating superfluids},
	volume = {59},
	issn = {0034-6861},
	url = {https://link.aps.org/doi/10.1103/RevModPhys.59.87},
	doi = {10.1103/RevModPhys.59.87},
	language = {en},
	number = {1},
	urldate = {2020-04-06},
	journal = {Reviews of Modern Physics},
	author = {Sonin, E. B.},
	month = jan,
	year = {1987},
	pages = {87--155},
}

@book{feynmanChapterIIApplication1955,
	title = {Chapter {II} {Application} of {Quantum} {Mechanics} to {Liquid} {Helium}},
	volume = {1},
	isbn = {978-0-444-53307-4},
	url = {https://linkinghub.elsevier.com/retrieve/pii/S0079641708600773},
	urldate = {2020-07-18},
	booktitle = {Progress in {Low} {Temperature} {Physics}},
	publisher = {Elsevier},
	author = {Feynman, R.P.},
	year = {1955},
	doi = {10.1016/S0079-6417(08)60077-3},
	pages = {17--53},
}

@article{sheUniversalScalingLaws1994,
	title = {Universal scaling laws in fully developed turbulence},
	volume = {72},
	issn = {0031-9007},
	url = {https://link.aps.org/doi/10.1103/PhysRevLett.72.336},
	doi = {10.1103/PhysRevLett.72.336},
	language = {en},
	number = {3},
	urldate = {2020-08-27},
	journal = {Physical Review Letters},
	author = {She, Zhen-Su and Leveque, Emmanuel},
	month = jan,
	year = {1994},
	pages = {336--339},
}

@article{spalartAirplaneTraiingVortices1998,
	title = {Airplane traiing vortices},
	volume = {30},
	issn = {0066-4189, 1545-4479},
	url = {https://www.annualreviews.org/doi/10.1146/annurev.fluid.30.1.107},
	doi = {10.1146/annurev.fluid.30.1.107},
	language = {en},
	number = {1},
	urldate = {2025-11-06},
	journal = {Annual Review of Fluid Mechanics},
	author = {Spalart, Philippe R.},
	month = jan,
	year = {1998},
	pages = {107--138},
}

@article{dariosSulMotoLiquido1906,
	title = {Sul moto d’un liquido indefinito con un filetto vorticoso di forma qualunque},
	volume = {22},
	issn = {0009-725X, 1973-4409},
	url = {http://link.springer.com/10.1007/BF03018608},
	doi = {10.1007/BF03018608},
	language = {it},
	number = {1},
	urldate = {2020-04-07},
	journal = {Rendiconti del Circolo Matematico di Palermo},
	author = {Da Rios, Luigi Sante},
	month = dec,
	year = {1906},
	pages = {117--135},
}

@book{saffmanVortexDynamics1993,
	edition = {1},
	title = {Vortex {Dynamics}},
	isbn = {978-0-521-47739-0 978-0-521-42058-7 978-0-511-62406-3},
	url = {https://www.cambridge.org/core/product/identifier/9780511624063/type/book},
	urldate = {2020-07-20},
	publisher = {Cambridge University Press},
	author = {Saffman, P. G.},
	month = jan,
	year = {1993},
	doi = {10.1017/CBO9780511624063},
}

@article{dahlCentrifugalWavesTornadoLike2021,
  title = {Centrifugal {{Waves}} in {{Tornado-Like Vortices}}: {{Kelvin}}'s {{Solutions}} and {{Their Applications}} to {{Multiple-Vortex Development}} and {{Vortex Breakdown}}},
  shorttitle = {Centrifugal {{Waves}} in {{Tornado-Like Vortices}}},
  author = {Dahl, Johannes M. L.},
  year = 2021,
  month = oct,
  journal = {Monthly Weather Review},
  volume = {149},
  number = {10},
  pages = {3173--3216},
  issn = {0027-0644, 1520-0493},
  doi = {10.1175/MWR-D-20-0426.1},
  urldate = {2025-11-06},
  copyright = {http://www.ametsoc.org/PUBSReuseLicenses}
}

@article{fabreKelvinWavesSingular2006,
  title = {Kelvin Waves and the Singular Modes of the {{Lamb}}--{{Oseen}} Vortex},
  author = {Fabre, David and Sipp, Denis and Jacquin, Laurent},
  year = 2006,
  month = mar,
  journal = {J. Fluid Mech.},
  volume = {551},
  pages = {235--274},
  issn = {0022-1120, 1469-7645},
  doi = {10.1017/S0022112005008463},
  urldate = {2025-11-06},
  copyright = {https://www.cambridge.org/core/terms},
  langid = {english}
}

@article{fondaDirectObservationKelvin2014,
  title = {Direct Observation of {{Kelvin}} Waves Excited by Quantized Vortex Reconnection},
  author = {Fonda, E. and Meichle, D. P. and Ouellette, N. T. and Hormoz, S. and Lathrop, D. P.},
  year = 2014,
  month = mar,
  journal = {Proceedings of the National Academy of Sciences},
  volume = {111},
  number = {Supplement1},
  pages = {4707--4710},
  issn = {0027-8424, 1091-6490},
  doi = {10.1073/pnas.1312536110},
  urldate = {2020-04-02},
  langid = {english}
}

@article{hasimotoSolitonVortexFilament1972,
  title = {A Soliton on a Vortex Filament},
  author = {Hasimoto, Hidenori},
  year = 1972,
  month = feb,
  journal = {J. Fluid Mech.},
  volume = {51},
  number = {3},
  pages = {477--485},
  issn = {0022-1120, 1469-7645},
  doi = {10.1017/S0022112072002307},
  urldate = {2020-04-06},
  langid = {english}
}

@article{krstulovicKelvinwaveCascadeDissipation2012,
  title = {Kelvin-Wave Cascade and Dissipation in Low-Temperature Superfluid Vortices},
  author = {Krstulovic, Giorgio},
  year = 2012,
  month = nov,
  journal = {Phys. Rev. E},
  volume = {86},
  number = {5},
  pages = {055301},
  issn = {1539-3755, 1550-2376},
  doi = {10.1103/PhysRevE.86.055301},
  urldate = {2020-03-01},
  copyright = {All rights reserved},
  langid = {english}
}

@article{ledizesAsymptoticDescriptionVortex2005,
  title = {An Asymptotic Description of Vortex {{Kelvin}} Modes},
  author = {Le Diz{\'e}s, St{\'e}phane and Lacaze, Laurent},
  year = 2005,
  month = nov,
  journal = {J. Fluid Mech.},
  volume = {542},
  pages = {69--96},
  issn = {0022-1120, 1469-7645},
  doi = {10.1017/S0022112005005185},
  urldate = {2025-11-06},
  copyright = {https://www.cambridge.org/core/terms},
  langid = {english}
}

@article{lvovWeakTurbulenceKelvin2010,
  title = {Weak Turbulence of {{Kelvin}} Waves in Superfluid {{He}}},
  author = {L'vov, Victor S. and Nazarenko, Sergey},
  year = 2010,
  month = aug,
  journal = {Low Temperature Physics},
  volume = {36},
  number = {8},
  pages = {785--791},
  issn = {1063-777X, 1090-6517},
  doi = {10.1063/1.3499242},
  urldate = {2020-02-07},
  langid = {english}
}

@article{minowaDirectExcitationKelvin2025,
  title = {Direct Excitation of {{Kelvin}} Waves on Quantized Vortices},
  author = {Minowa, Yosuke and Yasui, Yuki and Nakagawa, Tomo and Inui, Sosuke and Tsubota, Makoto and Ashida, Masaaki},
  year = 2025,
  month = feb,
  journal = {Nat. Phys.},
  volume = {21},
  number = {2},
  pages = {233--238},
  issn = {1745-2473, 1745-2481},
  doi = {10.1038/s41567-024-02720-9},
  urldate = {2025-10-21},
  langid = {english}
}

@article{mullerKolmogorovKelvinWave2020,
  title = {Kolmogorov and {{Kelvin}} Wave Cascades in a Generalized Model for Quantum Turbulence},
  author = {M{\"u}ller, Nicol{\'a}s P. and Krstulovic, Giorgio},
  year = 2020,
  month = oct,
  journal = {Phys. Rev. B},
  volume = {102},
  number = {13},
  pages = {134513},
  issn = {2469-9950, 2469-9969},
  doi = {10.1103/PhysRevB.102.134513},
  urldate = {2020-11-10},
  copyright = {All rights reserved},
  langid = {english}
}

\end{document}